\documentclass[aps, prd, preprint, amsfonts,
  amssymb, amsmath, showpacs, a4paper,nofootinbib,superscriptaddress]{revtex4-1}  
\usepackage[T1]{fontenc} 

\usepackage{slashed}
\usepackage{braket}

\newcommand{\be}{\begin{equation}}
\newcommand{\ee}{\end{equation}}
\newcommand{\bea}{\begin{eqnarray}}
\newcommand{\eea}{\end{eqnarray}}

\newcommand{\nn}{\nonumber\\}

\usepackage{color}
\usepackage{amsmath,amssymb,amscd}
\usepackage{listings}
\usepackage{dsfont}
\usepackage{slashed}
\usepackage{color}
\usepackage[normalem]{ulem}

\usepackage[pdftex]{graphicx}
\usepackage{epstopdf}
\usepackage{subfigure}
\usepackage{epsfig}
\usepackage{listings}
\usepackage{hyperref}

\begin{document} 

\title{Spontaneously breaking Non-Abelian Gauge Symmetry in \\ Non-Hermitian Field Theories}

\author{Jean Alexandre}
\email{jean.alexandre@kcl.ac.uk}
\affiliation{Department of Physics, King's College London,\\ 
London WC2R 2LS, United Kingdom}

\author{John Ellis}
\email{john.ellis@cern.ch}
\affiliation{Department of Physics, King's College London,\\ 
London WC2R 2LS, United Kingdom}
\affiliation{National Institute of Chemical Physics \& Biophysics, R\"avala 10, 10143 Tallinn, Estonia}
\affiliation{Theoretical Physics Department, CERN, CH-1211 Geneva 23, Switzerland}

\author{Peter Millington}
\email{p.millington@nottingham.ac.uk}
\affiliation{School of Physics and Astronomy, University of Nottingham,\\ Nottingham NG7 2RD, United Kingdom\vspace{1em}}

\author{Dries Seynaeve}
\email{dries.seynaeve@kcl.ac.uk}
\affiliation{Department of Physics, King's College London,\\ 
London WC2R 2LS, United Kingdom}

\begin{abstract}

We generalise our previous formulation of gauge-invariant $\mathcal{PT}$-symmetric field theories
to include models with non-Abelian symmetries and discuss the extension to such models of the 
Englert-Brout-Higgs-Kibble mechanism for generating masses for vector bosons. 
As in the Abelian case, the non-Abelian gauge fields are coupled to non-conserved currents.
We present a consistent scheme for gauge fixing, demonstrating Becchi-Rouet-Stora-Tyutin invariance, and show
that the particle spectrum and interactions are gauge invariant. We exhibit the masses that gauge bosons 
in the simplest two-doublet SU(2)$\times$U(1) model acquire when certain scalar fields develop
vacuum expectation values: they and scalar masses depend quartically on the non-Hermitian mass parameter $\mu$. 
The bosonic mass spectrum differs substantially from that in a Hermitian two-doublet model.
This non-Hermitian extension of the Standard Model opens a new direction for particle model building,
with distinctive predictions to be explored further.\\
~~\\
KCL-PH-TH/2019-76, CERN-TH-2019-161
~~\\
February 2020
\\
\noindent\footnotesize{This is an author-prepared post-print of \href{https://doi.org/10.1103/PhysRevD.101.035008}{Phys.\ Rev.\ D {\bf 101} (2020) 035008}, published by the American Physical Society under the terms of the \href{https://creativecommons.org/licenses/by/4.0/}{CC BY 4.0} license (funded by SCOAP\textsuperscript{3}).}
\end{abstract}

\maketitle


\section{Introduction}

Recent years have seen increasing interest in quantum-mechanical models with non-Hermitian, $\mathcal{PT}$-symmetric
Hamiltonians~\cite{Bender:1998ke,Bender:2002vv,Bender:2005tb}, which have been shown to possess real energy spectra that are bounded below, and have
extensive applications in photonics and other fields~\cite{Longhi,El-Ganainy,Ashida}. This interest has extended to $\mathcal{PT}$-symmetric
quantum field theories with non-Hermitian Lagrangians, such as a scalar field theory with an $i \phi^3$ interaction~
\cite{Blencowe:1997sy, Bender:2004vn, Jones:2004gp, Bender:2013qp},
which has been shown to possess a physically meaningful effective potential, a $\mathcal{PT}$-symmetric $-\phi^4$
scalar field theory~\cite{Shalaby:2009xda}, and a fermionic model with a non-Hermitian mass term $\propto \overline{\psi} \gamma_5 \psi$
that is unitary and has a conserved current~\cite{Bender:2005hf, Alexandre:2015oha}.
Such non-Hermitian quantum field theories have been applied to describe neutrino masses and oscillations~
\cite{JonesSmith:2009wy, Alexandre:2015kra, Alexandre:2017fpq, Ohlsson:2015xsa} (see also a similar lattice fermion model~\cite{Chernodub:2017lmx}),
and have also been considered in connection with dark matter~\cite{Rodionov:2017dqt} and decays of the Higgs boson~\cite{Korchin:2016rsf}. We note also
that effective non-Hermitian Hamiltonians can also be used to describe unstable systems with particle mixing~\cite{Pilaftsis:1997dr}.

The formulation of $\mathcal{PT}$-symmetric quantum field theories was 
extended in Refs.~\cite{Alexandre:2015kra, alexandre2017symmetries} to include
an Abelian gauge symmetry. A particularity of this formulation is that the gauge field is coupled to a non-conserved
current.
The next step was to study spontaneous
symmetry breaking and the Goldstone theorem~\cite{GSW1, GSW2, GSW3} in a {non-Hermitian,} $\mathcal{PT}$-symmetric quantum field theory, which was done
in Ref.~\cite{AEMS1} {(cf.~the alternative approach of Refs.~\cite{Mannheim:2018dur, Fring:2019hue})}, where we exhibited a specific example with two complex scalar 
fields and a non-Hermitian bilinear scalar coupling $\mu^2$,
in which there is a massless boson at both the tree and one-loop levels
~\footnote{The behaviours of Goldstone modes in different phases of $\mathcal{PT}$ symmetry have also been studied in Ref.~\cite{Fring:2019hue}.}. 
We note that physical observables depend only on $\mu^4$ and are therefore independent of the ambiguity in the sign of $\mu^2$ 
that arises from the non-Hermiticity of the model.
We subsequently explored in Ref.~\cite{AEMS2} 
the $\mathcal{PT}$-symmetric extension of the Englert-Brout-Higgs mechanism~\cite{Englert:1964et, Higgs:1964pj} for generating a mass for the 
Abelian gauge boson in a manner consistent with renormalisability of the quantum field theory. For summary of these works, see Ref.~\cite{Millington:2019dfn}.

In this paper, we further develop the formulation of $\mathcal{PT}$-symmetric gauge theories to include a
non-Abelian gauge symmetry and Kibble's non-Abelian generalisation~\cite{Kibble} of the Englert-Brout-Higgs mechanism.
We study a minimal extension of the model considered in Refs.~\cite{AEMS1,AEMS2} that contains two complex
scalar doublets and admits the same SU(2)$\times$U(1) gauge symmetry as the Standard Model.
We show how the gauge can be fixed in a consistent manner
and demonstrate Becchi-Rouet-Stora-Tyutin (BRST) invariance~\cite{BRST}. We explore the scalar vacuum expectation values (vev's) and
tree-level spectra of the gauge and scalar boson masses in a simple version of the model with a 
single quartic coupling. They depend quartically on the non-Hermitian coupling $\mu^2$ and
differ significantly from the masses in the conventional Hermitian two-Higgs-doublet model (2HDM, see Ref.~\cite{Branco:2011iw}). 
Thus, this non-Hermitian
extension of the Standard Model offers prospects for distinctive experimental predictions that may be explored 
further in a systematic programme of $\mathcal{PT}$-symmetric phenomenology.

\section{Scalar Lagrangian}

In this Section, we extend the non-Hermitian model of Ref.~\cite{alexandre2017symmetries} to include 
two complex scalar doublets, giving the non-Hermitian 2HDM on which we base the discussion of non-Abelian gauge symmetry and its
breaking in the next Section.

\subsection{Lagrangian}

We follow here similar steps to those described in Ref.~\cite{alexandre2017symmetries}, starting with the Lagrangian   
\bea \label{eq:ScalarLagrangian}
\mathcal{L} &=& \partial_\alpha \Phi_1^\dagger \partial^\alpha \Phi_1 + \partial_\alpha \Phi_2^\dagger \partial^\alpha \Phi_2 - m_1^2 |\Phi_1|^2 
- m_2^2 |\Phi_2|^2 \nn &&- \mu^2 \left( \Phi_1^\dagger \Phi_2 - \Phi_2^\dagger \Phi_1 \right) - \frac{\kappa}{4}|\Phi_1|^4~,
\eea
where $\Phi_i$ are complex doublets
\be
\Phi_i = \begin{pmatrix}
\phi_{ia} \\ \phi_{ib}
\end{pmatrix}~,\qquad i=1,2
\ee
and $\mu$ is a non-Hermitian mass parameter.
This system is invariant under the $\mathcal{PT}$-symmetry, acting on the $c$-number fields as
\bea \label{eq:PTtransf}
\mathcal{PT}:\qquad \Phi_1(t, x) &\rightarrow& \Phi_1'(-t, -x) = \Phi_1^*(t, x)~, \nn
\Phi_2(t, x) &\rightarrow& \Phi_2'(-t, -x) = - \Phi_2^*(t, x)~,
\eea
under which $\Phi_1$ is a scalar doublet whereas $\Phi_2$ is a pseudoscalar doublet. The eigenvalues of the 
squared mass matrix
\be\label{eigenvalues}
M_\pm^2=\frac{m_1^2+m_2^2}{2} \; {\pm} \; \frac{1}{2}\sqrt{(m_1^2-m_2^2)^2-4\mu^4}
\ee
are real provided the following inequality holds:
\be
2|\mu^2|\le|m_1^2-m_2^2|~,
\ee
which is assumed throughout the first two Sections of this work. Note that the eigenvalues become degenerate at $|\mu^2|=|m_1^2-m_2^2|/2$. 
This marks the \emph{exceptional point}, which lies at the boundary between the regions of unbroken and broken $\mathcal{PT}$ symmetry. 
At this point, the squared mass matrix becomes defective and we lose an eigenvector. We discuss these exceptional points further 
in Section~\ref{sec:exceptionalpoints}.

Because of the non-Hermitian mass term proportional to $\mu^2$, the equations of motion one obtains by varying the action with respect 
to $\Phi_i$ or to $\Phi_i^\dagger$ are not equivalent for non-trivial solutions, i.e.
\be
\label{eq:variations}
\frac{\delta S}{\delta \Phi_i^\dagger} \equiv 
\frac{\partial \mathcal{L}}{\partial \Phi_i^\dagger} - \partial_\alpha \frac{\partial \mathcal{L}}{\partial \big( \partial_\alpha \Phi_i^\dagger \big)} =0
\qquad\nLeftrightarrow\qquad \frac{\delta S}{\delta \Phi_i} \equiv 
\frac{\partial \mathcal{L}}{\partial \Phi_i} - \partial_\alpha \frac{\partial \mathcal{L}}{\partial \big( \partial_\alpha \Phi_i \big)}=0~.
\ee
These two sets of equations of motion are related by $\mathcal{PT}$-symmetry though or, equivalently, by a change in the sign of $\mu^2$. 
As can be seen from the eigenvalues (\ref{eigenvalues}), observables depend on $\mu^4$ only, so these two sets of equations of motion are physically equivalent. 
This is also valid at the quantum level, see Ref.~\cite{AEMS2}, as can be derived from the reality of the partition function, 
provided the sources for the scalar fields 
satisfy appropriate $\mathcal{PT}$ properties.

We choose here the equations of motion provided by the variation of the action with respect to $\Phi_i^\dagger$: 
\begin{subequations}
\label{eq:Scalareom}
\bea
0 &=& \Box \Phi_1 + m_1^2 \Phi_1 + \mu^2 \Phi_2 + \frac{\kappa}{2} |\Phi_1|^2 \Phi_1 ~,\\
0 &=& \Box \Phi_2 + m_2^2 \Phi_2 - \mu^2 \Phi_1~,
\eea
\end{subequations}
together with their Hermitian conjugates
\begin{subequations}
\bea
0 &=& \Box \Phi_1^\dagger + m_1^2 \Phi_1^\dagger + \mu^2 \Phi_2^\dagger + \frac{\kappa}{2} |\Phi_1|^2 \Phi_1^\dagger ~,\\
0 &=& \Box \Phi_2^\dagger + m_2^2 \Phi_2^\dagger - \mu^2 \Phi_1^\dagger~.
\eea
\end{subequations}
We note that this formulation differs from that suggested in Ref.~\cite{Mannheim:2018dur},
where the author introduces a similarity transformation that transforms the non-Hermitian Lagrangian $\mathcal{L}$ to a Hermitian one $\mathcal{L}'$. 
The difference in approach is reflected in differences in the masses of the gauge fields, which
 we discuss in Section~\ref{sec:exceptionalpoints}.

\subsection{Conserved currents}

The Lagrangian (\ref{eq:ScalarLagrangian}) is invariant under the U(1) transformations
\begin{subequations}
\bea
\Phi_1 &\rightarrow& e^{-i \frac{g'}{2} \beta_0} \Phi_1~,\\
\Phi_2 &\rightarrow& e^{-i \frac{g'}{2} \beta_0} \Phi_2~,
\eea
\end{subequations}
which correspond to the current
\be \label{I+}
I^\alpha_+ = i \frac{g'}{2} \left(  \left[ \Phi_1^\dagger \Big( \partial^\alpha \Phi_1 \Big) - \Big( \partial^\alpha \Phi_1^\dagger \Big) \Phi_1 \right]  
+ \left[ \Phi_2^\dagger \Big( \partial^\alpha \Phi_2 \Big) - \Big( \partial^\alpha \Phi_2^\dagger \Big) \Phi_2 \right] \right)~,
\ee
and also invariant under the SU(2) transformations
\begin{subequations}
\bea
\Phi_1 &\rightarrow& e^{-i\frac{g}{2} \vec{\beta} \cdot \vec{\tau}} \Phi_1~,\\
\Phi_2 &\rightarrow& e^{-i\frac{g}{2} \vec{\beta} \cdot \vec{\tau}} \Phi_2~,
\eea
\end{subequations}
which correspond to the current
\be\label{J+}
\vec{J}^\alpha_+ = i \frac{g}{2} \left(  \left[ \Phi_1^\dagger \vec{\tau} \Big( \partial^\alpha \Phi_1 \Big) 
- \Big( \partial^\alpha \Phi_1^\dagger \Big) \vec{\tau} \Phi_1 \right]  + \left[ \Phi_2^\dagger \vec{\tau} \Big( \partial^\alpha \Phi_2 \Big) 
- \Big( \partial^\alpha \Phi_2^\dagger \Big) \vec{\tau} \Phi_2 \right] \right)~,
\ee
where $\vec{\tau} = ( \tau_1, \tau_2, \tau_3)$ is composed of the Pauli matrices.

The equations of motion (\ref{eq:Scalareom}) show, however, that these currents are not conserved:
\begin{subequations}
\label{divergencecurrents}
\bea
\partial_\alpha I_+^\alpha &=& ig' \mu^2  \left( \Phi_2^\dagger \Phi_1 - \Phi_1^\dagger \Phi_2 \right) ~,\\
\partial_\alpha \vec{J}^\alpha_+ &=& ig\mu^2 \left( \Phi_2^\dagger \vec{\tau} \Phi_1 - \Phi_1^\dagger \vec{\tau} \Phi_2 \right)~,
\eea
\end{subequations}
except at the Hermitian point $\mu^2=0$. The fact that symmetries of the Lagrangian do not correspond to conserved currents for non-Hermitian 
theories is a direct consequence of the fact that the two functional variations in Eq.~\eqref{eq:variations} cannot vanish simultaneously for non-trivial solutions.  
Instead, a careful treatment of Noether's original derivation~\cite{Noether} shows that there still exist conserved currents for non-Hermitian theories, 
but these correspond to transformations that do not leave the Lagrangian invariant~\cite{alexandre2017symmetries} (see also Ref.~\cite{Alexandre:2017erl} for a summary).

In the present model, we find that the conserved currents are, in fact,
\begin{subequations}
\bea
I_-^\alpha &=& i \frac{g'}{2} \left(  \left[ \Phi_1^\dagger \Big( \partial^\alpha \Phi_1 \Big) - \Big( \partial^\alpha \Phi_1^\dagger \Big) \Phi_1 \right]  
- \left[ \Phi_2^\dagger \Big( \partial^\alpha \Phi_2 \Big) - \Big( \partial^\alpha \Phi_2^\dagger \Big) \Phi_2 \right] \right)~,\\
\vec{J}^\alpha_- &=& i \frac{g}{2} \left(  \left[ \Phi_1^\dagger \vec{\tau} \Big( \partial^\alpha \Phi_1 \Big) 
- \Big( \partial^\alpha \Phi_1^\dagger \Big) \vec{\tau} \Phi_1 \right]  - \left[ \Phi_2^\dagger \vec{\tau} \Big( \partial^\alpha \Phi_2 \Big) 
- \Big( \partial^\alpha \Phi_2^\dagger \Big) \vec{\tau} \Phi_2 \right] \right)~,
\eea
\end{subequations}
which correspond to the following transformations:
\begin{subequations}
\bea
\Phi_1 &\rightarrow& e^{-i \frac{g'}{2}\beta_0} \Phi_1~,\\
\Phi_2 &\rightarrow& e^{+i \frac{g'}{2}\beta_0} \Phi_2~,
\eea
\end{subequations}
and
\begin{subequations}
\bea
\Phi_1 &\rightarrow& e^{-i \frac{g}{2}\vec{\beta}\cdot \vec{\tau}} \Phi_1~,\\
\Phi_2 &\rightarrow& e^{+i \frac{g}{2}\vec{\beta}\cdot \vec{\tau}} \Phi_2~.
\eea
\end{subequations}
The relative sign between the charge assignments of the two fields reflects the usual interpretation of viable $\mathcal{PT}$-symmetric 
theories as systems with coupled gain and loss.

\section{Gauging the Scalar Model}

Since the conserved currents do not correspond to the usual Noether currents, gauging the 
model (\ref{eq:ScalarLagrangian}) is non-trivial, as we describe in this Section,
generalising the approach taken in Ref.~\cite{AEMS2} to the non-Abelian case. {We refer to the \emph{non-conserved} currents corresponding to 
symmetries of the Lagrangian as \emph{Noether currents}, but note that the \emph{conserved} currents are in fact those consistent with Noether's 
original derivation (see Ref.~\cite{alexandre2017symmetries}).}

\subsection{Coupling to the Noether currents}

We introduce an Abelian gauge field $B^\alpha$ and an $SU(2)$ gauge field $\vec W^\alpha$, together with the SU(2)$\times$U(1) gauge transformations
\begin{subequations}
\bea
\Phi_i &\rightarrow& e^{-i \frac{g'}{2} \beta_0} e^{-i \frac{g}{2} \vec{\beta} \cdot\vec{\tau}} \Phi_i ~,\\
\vec{W}^\alpha &\rightarrow& \vec{W}^\alpha + g \left( \vec{\beta} \times \vec{W}^\alpha \right) + \partial^\alpha \vec{\beta} = \vec{W}^\alpha 
+ \mathcal{D}^\alpha \vec{\beta}~,\\
B^\alpha &\rightarrow& B^\alpha + \partial^\alpha \beta_0~,
\eea
\end{subequations}
where $\mathcal{D}^\alpha \vec{\beta} = \partial^\alpha \vec{\beta} - g (\vec{W}^\alpha \times \vec{\beta})$.
In order to write a gauge-invariant theory, one should couple the gauge fields to the Noether currents, such that the scalar kinetic terms are given by
\bea\label{LK}
\mathcal{L}_{\rm kin} &=& \left[ D_\alpha \Phi_1 \right]^\dagger D^\alpha \Phi_1 + \left[ D_\alpha \Phi_2 \right]^\dagger D^\alpha \Phi_2 \nn
&=& \partial_\alpha \Phi_1^\dagger \partial^\alpha \Phi_1 + \partial_\alpha \Phi_2^\dagger \partial^\alpha \Phi_2 + \frac{i}{2} \partial_\alpha \Phi_1^\dagger \left( g' B^\alpha + g \vec{\tau} \cdot \vec{W}^\alpha \right) \Phi_1 \, , \nn
&&- \frac{i}{2} \Phi_1^\dagger \left( g' B^\alpha + g \vec{\tau} \cdot \vec{W}^\alpha \right) \partial_\alpha \Phi_1 + \frac{i}{2} \partial_\alpha \Phi_2^\dagger \left( g' B^\alpha + g \vec{\tau} \cdot \vec{W}^\alpha \right) \Phi_2  \, , \nn
&&- \frac{i}{2} \Phi_2^\dagger \left( g' B^\alpha + g \vec{\tau} \cdot \vec{W}^\alpha \right) \partial_\alpha \Phi_2 + \frac{1}{4} \Phi_1^\dagger \left( g' B + g \vec{\tau} \cdot \vec{W} \right)^2 \Phi_1 \nn
&&+ \frac{1}{4} \Phi_2^\dagger \left( g' B + g \vec{\tau} \cdot \vec{W} \right)^2 \Phi_2~, 
\eea
where $D_\alpha$ is given by the usual minimal-coupling prescription, i.e.,
\be
D^\alpha \Phi_i = \partial^\alpha \Phi_i + \frac{ig'}{2} B^\alpha \Phi_i + \frac{ig}{2} \left[ \vec{\tau} \cdot \vec{W}^\alpha \right] \Phi_i~.
\ee
As in the Standard Model, we rotate the gauge fields as
\begin{subequations}  
\label{eq:newGaugefields}
\begin{gather}
B^\alpha = \cos \theta_{\rm W} A^\alpha - \sin \theta_{\rm W}Z^\alpha~,\\
W^\alpha_1 = \frac{W^\alpha + W^{\alpha \dagger}}{\sqrt{2}}~,\qquad
W^\alpha_3 = \sin \theta_{\rm W}A^\alpha + \cos \theta_{\rm W} Z^\alpha~,\qquad W^\alpha_2 = i \frac{W^\alpha - W^{\alpha \dagger}}{\sqrt{2}}~,
\end{gather}
\end{subequations}
where $\theta_{\rm W}$ is the weak mixing angle, to obtain
\bea \label{eq:KineticA}
\mathcal{L}_{\rm kin}  &=& \partial_\alpha \Phi_1^\dagger \partial^\alpha \Phi_1 + \partial_\alpha \Phi_2^\dagger \partial^\alpha \Phi_2  - W_\alpha \left[ \frac{J_{+,1}^\alpha + i J_{+,2}^\alpha}{\sqrt{2}} \right] - W_\alpha^\dagger \left[ \frac{J_{+,1}^\alpha - i J_{+,2}^\alpha}{\sqrt{2}} \right] \nn
&&- Z_\alpha \left[ J_{+,3}^\alpha \cos \theta_{\rm W} - I_+^\alpha \sin \theta_{\rm W} \right] - A_\alpha \left[ J_{+,3}^\alpha \sin \theta_{\rm W} + I_+^\alpha \cos \theta_{\rm W} \right]  \nn
&& + \frac{g^2}{2} W_\alpha^\dagger W^\alpha \left( |\Phi_1|^2 + |\Phi_2|^2\right) \nn
&&+ \frac{1}{4} Z_\alpha Z^\alpha \sum_i \Phi_i^\dagger \Big( \left[ g'^2 \sin^2 \theta_{\rm W} + g^2 \cos^2 \theta_{\rm W} \right] \mathbb{I} - 2 gg' \cos \theta_{\rm W} \sin \theta_{\rm W} \tau_3 \Big)\Phi_i
\nn
&&+ \frac{1}{4} A_\alpha A^\alpha \sum_i \Phi_i^\dagger \Big( \left[ g'^2 \cos^2 \theta_{\rm W} + g^2 \sin^2 \theta_{\rm W} \right] \mathbb{I} + 2 gg' \cos \theta_{\rm W} \sin \theta_{\rm W} \tau_3 \Big)\Phi_i \nn
&& + \frac{1}{4} Z_\alpha A^\alpha \sum_i \Phi_i^\dagger \Big( \left[ (g^2 - g'^2) \sin 2 \theta_{\rm W} \mathbb{I} + 2 g g' \cos 2 \theta_{\rm W} \tau_3 \right]\Big) \Phi_i \nn
&&+ \frac{1}{2} gg' \Big( \cos \theta_{\rm W} A^\alpha - \sin \theta_{\rm W} Z^\alpha \Big) \sum_i \Phi_i^\dagger \Big( W_\alpha \left[ \frac{\tau_1 + i \tau_2}{\sqrt{2}} \right] + W_\alpha^\dagger \left[ \frac{\tau_1 - i \tau_2}{\sqrt{2}} \right] \Big) \Phi_i~.\nonumber\\
\eea
Also as in the Standard Model, the Lagrangian for the gauge fields is
\bea\label{Lgauge}
\mathcal{L}_{\rm gauge} &=& -\frac{1}{4} \vec{W'}_{\alpha \beta} \cdot \vec{W'}^{\alpha \beta} -\frac{1}{4} B_{\alpha \beta} B^{\alpha \beta} \nn
&=& -\frac{1}{4} F_{\alpha \beta} F^{\alpha \beta} - \frac{1}{4} Z_{\alpha \beta} Z^{\alpha \beta} - \frac{1}{2}W_{\alpha \beta}^\dagger W^{\alpha \beta}~,
\eea
with
\begin{subequations}
\bea
\vec{W'}_{\alpha \beta} &=&  \partial_\beta \vec{W}_\alpha - \partial_\alpha \vec{W}_\beta + g \left( \vec{W}_\alpha \times \vec{W}_\beta \right)~,\\
B_{\alpha \beta} &=& \partial_\beta B_\alpha - \partial_\alpha B_\beta~,\\
W_{\alpha \beta} &=& \left[ \partial_\beta + ig \left( \sin \theta_{\rm W} A_\beta + \cos \theta_{\rm W} Z_\beta \right) \right] W_\alpha\nonumber\\
&&- \left[ \partial_\alpha + ig \left( \sin \theta_{\rm W} A_\alpha + \cos \theta_{\rm W} Z_\alpha \right) \right] W_\beta~,\\
F_{\alpha \beta} &=& \partial_\beta A_\alpha - \partial_\alpha A_\beta + ig \sin \theta_{\rm W} \left[ W_\alpha^\dagger W_\beta - W_\beta^\dagger W_\alpha  \right]~,\\
Z_{\alpha \beta} &=& \partial_\beta Z_\alpha - \partial_\alpha Z_\beta + ig \cos \theta_{\rm W} \left[ W_\alpha^\dagger W_\beta - W_\beta^\dagger W_\alpha  \right]~.
\eea
\end{subequations}

\subsection{Consistent field equations}

Since the gauge fields are coupled to currents that are not conserved, additional terms need to be added to the Lagrangian in order to have consistent field equations \cite{AEMS2}.
For this, it is enough to consider the usual gauge-fixing terms, which must be added to the 
{\it classical} equations of motion in the non-Hermitian case (not just at the quantum level in order
to define the path integral, as in the Hermitian case).
The gauge-fixing terms in the Lagrangian involve ghost fields $\vec{\eta}$ and $\vec{\bar{\eta}}$, taking the form
\bea \label{eq:GFLagr1}
\mathcal{L}_{\rm GF} &=& \partial_\alpha \vec{\bar{\eta}} \cdot \left[ \mathcal{D}^\alpha \vec{\eta} \right] 
- \frac{1}{2 \xi} \left[ \left( \partial_\alpha B^\alpha \right)^2 + |\partial_\alpha \vec{W}^\alpha|^2 \right] \nn 
&=& \partial_\alpha \bar{\chi}^\dagger \Big( \left[ \partial^\alpha + ig \left( \sin \theta_{\rm W} A^\alpha + \cos \theta_{\rm W} Z^\alpha \right) \right] \chi 
- ig  W^\alpha \eta_3  \Big) \nn 
&&+ \partial_\alpha \bar{\chi} \Big( \left[ \partial^\alpha - ig \left( \sin \theta_{\rm W} A^\alpha + \cos \theta_{\rm W} Z^\alpha \right] \right) \chi^\dagger 
+ ig  W^{\alpha \dagger} \eta_3  \Big) \nn
&&+ \partial_\alpha \bar{\eta}_3 \left( \partial^\alpha \eta_3 + ig \left[ W^\alpha \chi^\dagger - W^{\alpha \dagger} \chi \right) \right]  \nn
&&- \frac{1}{2 \xi} \left[ \left( \partial_\alpha A^\alpha \right)^2 + \left( \partial_\alpha Z^\alpha \right)^2 + 2 |\partial_\alpha W^\alpha|^2 \right]~,
\eea
where 
\be 
\bar{\chi} \equiv \frac{\bar{\eta}_1 - i \bar{\eta}_2}{\sqrt{2}}~,\qquad \chi \equiv \frac{\eta_1 - i \eta_2}{\sqrt{2}}~.
\ee
The equations of motion for the full Lagrangian are then given by
\begin{subequations}
\bea 
0 &=& D_\alpha D^\alpha \Phi_1 + m_1^2 \Phi_1 + \mu^2 \Phi_2 + \frac{\kappa}{2}|\Phi_1|^2 \Phi_1~, \\
0 &=& D_\alpha D^\alpha \Phi_2 + m_2^2 \Phi_2 - \mu^2 \Phi_1~, \\ \label{eq:eomW}
0 &=& \mathcal{D}_\beta \vec{W'}^{\beta \alpha} + \vec{\mathcal{J}}^\alpha_+ - \frac{1}{\xi} \partial^\alpha \partial^\beta \vec{W}_\beta - g \left( \partial^\alpha \vec{\bar{\eta}} \times \vec{\eta} \right)~, \\ \label{eq:eomB}
0 &=& \partial_\beta B^{\beta \alpha} + \mathcal{I}_+^\alpha - \frac{1}{\xi} \partial^\alpha \partial^\beta B_\beta~, \\ 
0 &=& \partial_\alpha \mathcal{D}^\alpha \vec{\eta}~,\\
0 &=& \mathcal{D}_\alpha \partial^\alpha \vec{\bar{\eta}}~,\label{eq:eomGhost}
\eea
\end{subequations}
together with their Hermitian conjugates, where
\begin{subequations}
\bea
\mathcal{I}^\alpha_+& \equiv & i \frac{g'}{2} \left( \left[ \Phi_1^\dagger (D^\alpha \Phi_1) - \left( D^\alpha \Phi_1 \right)^\dagger \Phi_1 \right] 
+ \left[ \Phi_2^\dagger (D^\alpha \Phi_2) - \left( D^\alpha \Phi_2 \right)^\dagger \Phi_2 \right] \right)  ~,\\
\vec{\mathcal{J}}^\alpha_+& \equiv &i \frac{g}{2} \left( \left[ \Phi_1^\dagger \vec{\tau} (D^\alpha \Phi_1) - \left( D^\alpha \Phi_1 \right)^\dagger \vec{\tau} \Phi_1 \right] 
+ \left[ \Phi_2^\dagger \vec{\tau} (D^\alpha \Phi_2) - \left( D^\alpha \Phi_2 \right)^\dagger \vec{\tau} \Phi_2 \right] \right)~.
\eea
\end{subequations}

Taking into account the current divergences (\ref{divergencecurrents}), the derivatives of the above equations of motion lead to the constraints
\begin{subequations}
\label{eq:gaugeCond1}
\bea
\label{eq:gaugeCond1A}
\frac{1}{\xi} \mathcal{D}_\alpha \partial^\alpha \partial^\beta \vec{W}_\beta &=& ig \mu^2 \left( \Phi_2^\dagger \vec{\tau} \Phi_1 - \Phi_1^\dagger \vec{\tau} \Phi_2 \right) 
- g \partial^\alpha  \vec{\bar{\eta}}\times \mathcal{D}_\alpha \vec{\eta}~,\\
\frac{1}{\xi}\Box \partial^\beta B_\beta &=& i g'\mu^2  \left( \Phi_2^\dagger \Phi_1 - \Phi_1^\dagger \Phi_2 \right)~,
\eea
\end{subequations}
which must be satisfied in order for the field equations to be consistent. 
As explained in the next Subsection, BRST 
symmetry allows one to write the latter constraints independently of the ghost fields, as
\begin{subequations}
\label{constraints}
\bea
\frac{1}{\xi} \mathcal{D}_\alpha \partial^\alpha \partial^\beta \vec{W}_\beta &=& \frac{ig \mu^2}{2} \left( \Phi_2^\dagger \vec{\tau} \Phi_1 - \Phi_1^\dagger \vec{\tau} \Phi_2 \right)~,\\ 
\frac{1}{\xi}\Box \partial^\beta B_\beta &=& i g'\mu^2 \left( \Phi_2^\dagger \Phi_1 - \Phi_1^\dagger \Phi_2 \right)~.
\eea
\end{subequations}

We can summarise our approach as follows. In order to respect gauge invariance, we need to couple the gauge fields to the Noether currents. However, because these currents are not conserved, 
we need to introduce gauge-fixing terms, which restrict gauge invariance, but imply consistent field equations. The residual gauge invariance is enough to ensure that gauge fields 
remain massless in the absence of spontaneous symmetry breaking (SSB), and it is defined by the
gauge functions $\beta_0,\vec\beta$ satisfying
\begin{subequations}
\bea 
\partial_\alpha\mathcal{D}^\alpha\vec\beta &=&0~,\\
\Box \beta_0&=&0~.
\eea
\end{subequations}
We therefore obtain a consistent gauge theory with a non-Hermitian scalar sector, as in the Abelian case~\cite{AEMS2}.

\subsection{BRST Transformation}

In this Subsection, we derive the gauge constraint (\ref{constraints}) for $\vec{W}_\beta$ using the BRST transformation, which is 
a residual symmetry of the Lagrangian after gauge fixing. In order to define it, one can introduce an auxiliary field $\vec{T}$ to write the gauge-fixing Lagrangian 
(\ref{eq:GFLagr1}) in the alternative form
\be \label{eq:GFLagr2}
\mathcal{L}_{\rm GF} = \partial_\alpha \vec{\bar{\eta}} \cdot {\cal D}^\alpha \vec{\eta} + \frac{\xi}{2} |\vec{T}|^2 - \vec{T} \cdot \partial^\alpha \vec{W}_\alpha 
- \frac{1}{2 \xi} \left( \partial_\alpha B^\alpha \right)^2~,
\ee
and the original Lagrangian (\ref{eq:GFLagr1}) can be recovered after integrating out $\vec{T}$.
The BRST transformations are defined as 
\begin{subequations}
\label{eq:BRST}
\bea
\delta \phi_i &=& -i\frac{g}{2} \theta \left( \vec{\tau} \cdot \vec{\eta} \right) \phi_i~,\\
\delta \vec{W}^\alpha &=& \theta {\cal D}^\alpha \vec{\eta}~,\\
\delta B^\alpha &=& 0~,\\
\delta \vec{\bar{\eta}} &=& - \theta \vec{T}~,\\
\delta \vec{\eta} &=& \frac{g}{2} \theta \left( \vec{\eta} \times \vec{\eta} \right)~,\\
\delta \vec{T} &=& 0~,
\eea
\end{subequations}
where $\theta$ is an infinitesimal Grassmann parameter.
The gauge-invariant terms (\ref{LK}) and (\ref{Lgauge}) in the Lagrangian are invariant under the BRST transformation, and the gauge-fixing Lagrangian (\ref{eq:GFLagr2})
transforms as a total derivative, so the action is invariant under this BRST transformation.
Using the auxiliary field $\vec T$, the equation of motion (\ref{eq:eomW}) for the gauge field $\vec{W}^\alpha$ can be written in the form
\be 
0 = \mathcal{D}_\beta \vec{W'}^{\beta \alpha} + \vec{\mathcal{J}}^\alpha_+ - \partial^\alpha \vec{T} - g \left( \partial^\alpha \vec{\bar{\eta}} \times \vec{\eta} \right)~,
\ee  
and a covariant derivative leads to 
\be \label{eq:gaugeCond2}
\mathcal{D}_\alpha \partial^\alpha \vec{T} = i g\mu^2 \left( \Phi_2^\dagger \vec{\tau} \Phi_1 - \Phi_1^\dagger \vec{\tau} \Phi_2 \right) 
- g \partial^\alpha \vec{\bar{\eta}} \times \mathcal{D}_\alpha \vec{\eta}~.
\ee
A BRST transformation of Eq.~(\ref{eq:eomGhost}) leads then to the relation 
\be\label{eq:cons}
0 = \delta \left( \mathcal{D}_\alpha \partial^\alpha \vec{\bar{\eta}} \right) = - \theta \left( \mathcal{D}_\alpha \partial^\alpha \vec T
- g \partial_\alpha \vec{\bar{\eta}} \times \mathcal{D}^\alpha \vec{\eta} \right)~,
\ee
so that
\be
\mathcal{D}_\alpha \partial^\alpha \vec{T} = g \partial_\alpha \vec{\bar{\eta}} \times \mathcal{D}^\alpha \vec{\eta}~,
\ee
which, together with Eq.~(\ref{eq:gaugeCond2}), leads to
\be \label{eq:gaugeCond3}
\mathcal{D}_\alpha \partial^\alpha \vec{T} = \frac{i g \mu^2}{2} \left( \Phi_2^\dagger \vec{\tau} \Phi_1 - \Phi_1^\dagger \vec{\tau} \Phi_2 \right)~. 
\ee
Since, from the equations of motion for $\vec{T}$, one finds 
\be 
\vec{T} = \frac{1}{\xi} \partial_\alpha \vec W^\alpha~,
\ee
one finally obtains the expected constraint 
\be \label{eq:gaugeCond4}
\frac{1}{\xi} \mathcal{D}_\alpha \partial^\alpha \partial^\beta \vec{W}_\beta = \frac{i g \mu^2}{2} \left( \Phi_2^\dagger \vec{\tau} \Phi_1 - \Phi_1^\dagger \vec{\tau} \Phi_2 \right)~,
\ee
which, unlike Eq.~(\ref{eq:gaugeCond1A}), is independent of the ghost fields.

For further discussions of BRST (and anti-BRST) symmetries in the context of non-Hermitian field theories, see Ref.~\cite{Raval:2018kqg}.

\section{Spontaneous Symmetry breaking}

Spontaneous symmetry breaking (SSB) is possible if the sign of $m_1^2$ in the Lagrangian (\ref{eq:ScalarLagrangian}) is changed, 
and we study here the corresponding scalar vacuum expectation values and vector masses.

\subsection{Vacuum expectation value}
\label{sec:vevs}

With this change of sign, the Lagrangian (\ref{eq:ScalarLagrangian}) has a symmetry-breaking vacuum that is given by
\begin{subequations}
\label{eq:vacuumCond}
\bea
\frac{\kappa}{2} |\langle \Phi_1 \rangle|^2 &=& m_1^2 - \frac{\mu^4}{m_2^2}~,\\
\langle \Phi_2 \rangle &=& \frac{\mu^2}{m_2^2}\, \langle \Phi_1 \rangle~,
\eea
\end{subequations}
which is physical as long as
\be \label{eq:MuCond1}
m_1^2 m_2^2 > \mu^4~.
\ee
The vacuum is defined up to a SU(2)$\times$U(1) transformation, and it can be chosen so that
\be\label{vacuum}
\langle \Phi_1 \rangle = \begin{pmatrix} 0 \\ v_1 \end{pmatrix} \equiv V_1~,
\qquad \langle \Phi_2 \rangle = \begin{pmatrix} 0 \\ v_2 \end{pmatrix} \equiv V_2~,
\ee
with
\be
v_1 = \sqrt{\frac{2}{\kappa} \left( m_1^2 - \frac{\mu^4}{m_2^2} \right)}~,\qquad v_2 = \frac{\mu^2}{m_2^2} \sqrt{\frac{2}{\kappa} \left( m_1^2 - \frac{\mu^4}{m_2^2} \right)}~.
\ee
With this choice, the vacuum expectation value is unbroken by the transformation
\be \label{eq:unbrokentrans}
\langle \Phi_i \rangle \rightarrow e^{-i \frac{e}{2} \beta_0 \left( \mathbb{I} + \tau_3 \right)} \langle \Phi_i \rangle = \begin{pmatrix}
e^{-ie \beta_0} & 0 \\ 0 & 1
\end{pmatrix} \langle \Phi_i \rangle = \langle \Phi_i \rangle~,
\ee
such that the Abelian subgroup of SU(2)$\times$U(1) generated by $\sigma = \mathbb{I} + \tau_3$ remains unbroken. 
This subgroup corresponds to the electromagnetic interaction, with Noether current
\bea \label{eq:unbrokenCurrent}
Q^\alpha &=& \frac{ie}{2} \left[ \Phi_1^\dagger \sigma (\partial^\alpha \Phi_1) - (\partial_1^\alpha \Phi_1^\dagger) \sigma \Phi_1 \right] 
+ \frac{ie}{2} \left[ \Phi_2^\dagger \sigma (\partial^\alpha \Phi_2) - (\partial_2^\alpha \Phi_2^\dagger)\sigma \Phi_2 \right] \nn
&=& \frac{e}{g'} I^\alpha_+ + \frac{e}{g} J_{+,3}^\alpha~.
\eea
From Eq.~(\ref{eq:KineticA}), we see that the gauge field $A^\mu$ couples to the current $ I_+^\alpha \cos \theta_{\rm W} + J_{+,3}^\alpha \sin \theta_{\rm W}$, which can be identified with
the current (\ref{eq:unbrokenCurrent}) if
\be
e = g' \cos \theta_{\rm W} = g \sin \theta_{\rm W}~.
\ee
The ${\rm U}(1)_{\rm EM}$ charge is conserved at the tree level, although the Noether current is in general not conserved. 
Exploration of the possibility of charge non-conservation beyond the tree level lies beyond the scope of this paper. 
Its existence and observability would in principle depend upon the completion of the bosonic model considered here to include fermions, 
which is also a topic for future work.

We can then express the scalar Lagrangian in terms of fluctuations around the vacuum (\ref{vacuum}) as
\bea
\mathcal{L}_{\rm scal} &=& \partial_\alpha \hat{\Phi}_1^\dagger \partial^\alpha \hat{\Phi}_1 + \partial_\alpha \hat{\Phi}_2^\dagger \partial^\alpha \hat{\Phi}_2 
+ \frac{2 \mu^4}{m_2^2} \left( V_1^\dagger \hat{\Phi}_1 \right) - 2 m_2^2 \left( V_2^\dagger \hat{\Phi}_2 \right) \nn
&&-m_2^2 |\hat{\Phi}_2|^2 + \frac{\mu^4}{m_2^2} |\hat{\Phi}_1|^2 - \frac{\kappa}{4} \left( V_1^\dagger \hat{\Phi}_1 
+ \hat{\Phi}_1^\dagger V_1 \right)^2 - \mu^2 \left( \hat{\Phi}_1^\dagger \hat{\Phi}_2 - \hat{\Phi}_2^\dagger \hat{\Phi}_1 \right) \nn
&&-\frac{\kappa}{2} \left( V_1^\dagger \hat{\Phi}_1 + \hat{\Phi}_1^\dagger V_1 \right) |\hat{\Phi}_1|^2 - \frac{\kappa}{4}|\hat{\Phi}_1|^4~,
\eea
where
\begin{subequations}
\label{eq:fluctuation}
\bea
\Phi_i = \hat{\Phi}_i + V_i = \begin{pmatrix}
\phi_i^+ \\ v_i + \rho_i + i \psi_i
\end{pmatrix}~,\\ \Phi_i^* = \hat{\Phi}_i^* + V_i = \begin{pmatrix}
\phi_i^- \\ v_i + \rho_i - i \psi_i
\end{pmatrix}~.
\eea
\end{subequations}
We note that the terms linear in fluctuations are a consequence of the non-Hermitian nature of the system.
However, they do not play a role in the equations of motion 
$\delta S/\delta\hat\Phi_i^\dagger\equiv0$, since they depend on $\hat\Phi_i$ only.
These equations of motion are
\begin{subequations}
\bea
0 &=& \Box \hat{\Phi}_1 - \frac{\mu^4}{m_2^2} \hat{\Phi}_1 + \frac{\kappa}{2} \left( V_1^\dagger \hat{\Phi}_1 + \hat{\Phi}_1^\dagger V_1 \right) V_1 
+ \mu^2 \hat{\Phi}_2 \nn &&+ \frac{\kappa}{2}|\hat{\Phi}_1|^2 V_1 + \frac{\kappa}{2} \left( V_1^\dagger \hat{\Phi}_1 + \hat{\Phi}_1^\dagger V_1 \right) \hat{\Phi}_1 
+ \frac{\kappa}{2}|\hat{\Phi}_1|^2 \hat{\Phi}_1~,\\
0 &=& \Box \hat{\Phi}_2 + m_2^2 \hat{\Phi}_2 - \mu^2 \hat{\Phi}_1~.
\eea
\end{subequations}

The massless Goldstone modes consist of charged and neutral fields:
\begin{subequations}
\label{eq:ScalarMasses}
\bea
G^\pm &=& \frac{1}{\sqrt{v_1^2 - v_2^2}} \left(
v_1 \phi_{1}^{\pm} - v_2 \phi_{2}^{\pm} \right)~,\\
G &=& \frac{1}{\sqrt{v_1^2 - v_2^2}} \left( v_1 \psi_1 - v_2 \psi_2 \right)~.
\eea
\end{subequations}
The remaining fields consist of a charged field and three neutral fields. The charged fields are given by   
\be\label{eq:Hpm}
H^\pm = \frac{1}{\sqrt{v_1^2 - v_2^2}}
\left( v_2 \phi_{1}^{\pm} - v_1 \phi_{2}^{\pm}\right)~,
\ee
and one neutral field is given by
\be\label{D}
D = \frac{1}{\sqrt{v_1^2 - v_2^2}} \left( v_2 \psi_1 - v_1 \psi_2  \right)~,
\ee
with degenerate squared mass
\be \label{eq:ScalarMasses1}
M^2 =\frac{v_1^2-v_2^2}{v_1v_2}\mu^2 = m_2^2-\frac{\mu^4}{m_2^2}~.
\ee
Finally, we can express the last two neutral fields as 
\begin{subequations}
\label{H}
\bea
H &=& \rho_1 \cosh \alpha - \rho_2 \sinh \alpha~,\\
h &=& \rho_1 \sinh \alpha - \rho_2 \cosh \alpha~,
\eea
\end{subequations}
with masses 
\begin{subequations} \label{eq:ScalarMasses2}
\bea
M_h^2 &=& \frac{1}{2} \left( m_2^2 + 2m_1^2 - 3\mu^4 / m_2^2 - \sqrt{\left( 2m_1^2 - m_2^2 - 3\mu^4 / m_2^2 \right)^2 - 4 \mu^4} \right)
\nn &=&\left(v_1^2 - v_2^2 \right) \left[ \lambda - \frac{\hat{\lambda} \cosh \left( \beta - \alpha \right)}{\sinh \left( \beta - \alpha \right)} \right]~,\\
M_H^2 &=& \frac{1}{2} \left( m_2^2 + 2m_1^2 - 3\mu^4 / m_2^2 + \sqrt{\left( 2m_1^2 - m_2^2 - 3\mu^4 / m_2^2 \right)^2 - 4 \mu^4} \right) 
\nn &=& \left( v_1^2 - v_2^2 \right) \left[ \lambda - \frac{\hat{\lambda} \sinh \left( \beta - \alpha \right)}{\cosh \left( \beta - \alpha \right)} \right]~,
\eea
\end{subequations}
where 
\begin{subequations}
\label{alphabeta}
\bea
\tanh \alpha &=& \frac{-\mu^2}{\left( M_H^2 - m_2^2 \right)}~,\\
\tanh \beta &=& \frac{v_2}{v_1}~,
\eea
\end{subequations}
and 
\begin{subequations}
\bea
\lambda &=& \kappa\,\cosh^4 \beta ~,\\ \hat{\lambda} &=& \frac{\kappa}{2} \sinh 2 \beta\, \cosh^2  \beta ~.
\eea
\end{subequations}
It is not obvious that $M^2$ is positive or that $M_H^2$ and $M_h^2$ are real, and we derive the corresponding conditions on $\mu^2$ in the next Section.

The eigenvectors of non-Hermitian matrices are not orthogonal with respect to the Hermitian inner product
\be 
\langle \phi , \varphi \rangle = \int_x \phi^\dagger \varphi~.
\ee  
In the case of $\mathcal{PT}$-symmetric theories, however, the eigenmodes of the non-Hermitian Hamiltonian are orthogonal with respect to the $\mathcal{PT}$ inner product
\be 
\langle \phi , \varphi \rangle_\mathcal{PT} = \int_x \left( \phi^\mathcal{PT} \right)^\mathsf{T} \varphi~,
\ee
and we have normalised the fields $G^\pm$, $G$, $H^\pm$, $D$, $H$ and $h$ accordingly. These eigenmodes are non-trivial linear combinations of 
the scalar components of $\Phi_1$ and the pseudoscalar components of $\Phi_2$ and, as such, they cannot be eigenstates of $\mathcal{P}$. 
Instead, the $\mathcal{P}$ transformation relates the left and right eigenmodes, which are distinct for a non-Hermitian Hamiltonian.

We remark that the $\mathcal{PT}$ norm used for the modes $G$, $G^{\pm}$, $D$ and $H^{\pm}$ in Eqs.~(\ref{eq:ScalarMasses}), 
(\ref{eq:Hpm}) and (\ref{D}) diverges when $\mu^2=\pm m_2^2$ ($v_1^2=v_2^2$). At this point --- the \emph{zero exceptional point}
described in Ref.~\cite{Fring:2019hue} --- we lose three eigendirections: $D\propto G$ and $H^{\pm}\propto G^{\pm}$. On the other hand, when $\mu^2=\pm T_{H(h)}$, where
\be 
T_{H(h)} = \frac{m_2^2}{9} \left( 6m_1^2 - m_2^2 +(-) 2 \sqrt{2 m_2^2 \left( 3m_1^2 - m_2^2 \right)} \right),
\ee
$|\alpha| \to \infty$ and the $\mathcal{PT}$ norm of $h$ and $H$ in Eq.~(\ref{H}) diverges. In this case, we lose one eigendirection: $H\propto h$. 
We discuss these exceptional points further in Subsection~\ref{sec:exceptionalpoints}.

\subsection{Conditions on $\mu^2$}

Ensuring that we are in a physical regime of spontaneous symmetry breaking leads to a number of constraints on the parameter $\mu^2$:
\begin{enumerate}
\item[I] In order for the symmetry to be broken [see Eq.~(\ref{eq:vacuumCond})], we require that
\be \label{eq:mucondI}
\mu^4 < m_1^2 m_2^2~.
\ee

\item[II] In order to ensure that the squared mass $M^2$, defined in Eq.~(\ref{eq:ScalarMasses1}), remains positive, we require that
\be \label{eq:mucondII}
\mu^4 < m_2^4~.
\ee

\item[III] In order for the squared masses $M_h^2$ and $M_H^2$, defined in Eq.~(\ref{eq:ScalarMasses2}), to be real, we require that
\be \label{eq:mucondIII}
4 \mu^4 \leq \left( 2m_1^2 - m_2^2 - \frac{3\mu^4}{m_2^2} \right)^2~. 
\ee
We remark that in the region $4 \mu^4 \geq \left( 2m_1^2 - m_2^2 - \frac{3\mu^4}{m_2^2} \right)^2$ the squared mass matrix cannot be brought to a Hermitian form by a similarity 
transformation~\cite{Mannheim:2018dur}.

\end{enumerate}

\begin{figure}[!h]
  \includegraphics[width=0.9\linewidth]{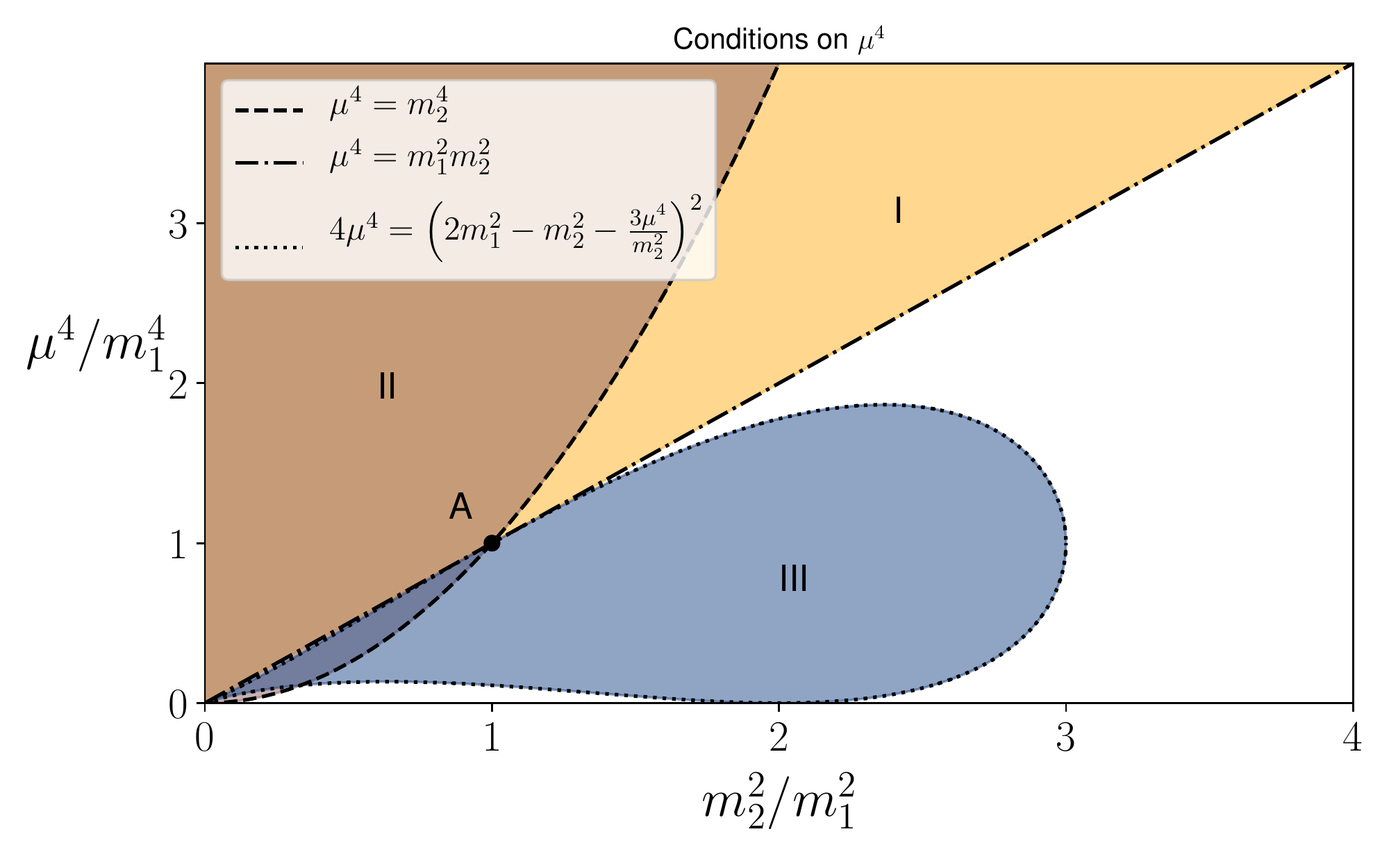}
  \caption{{\it The excluded regions for the parameter $\mu^4$, corresponding to the constraints I, II and III, plotted as functions of $m_2^2/m_1^2$. 
  Region I corresponds to the symmetric phase of the SU(2)$\times$U(1) symmetry [see Eq.~(\ref{eq:mucondI})]. Region II corresponds to the broken phase 
  of $\mathcal{PT}$ symmetry [see Eq.~(\ref{eq:mucondII})] in which $M^2$ is negative. Region III corresponds to the broken phase of $\mathcal{PT}$ 
  symmetry in which $M_h^2$ and $M_H^2$ [see Eq.~(\ref{eq:mucondIII})] are complex.
  The unshaded region corresponds to a physical SSB phase for the SU(2)$\times$U(1) symmetry. For $m_2^2/m_1^2 < 1/3$, the allowed region is determined only by condition II. 
  For $m_1^2/3 < m_2^2 < 3m_1^2$, the allowed region is determined by conditions I and III. Lastly, in the region $m_2^2 > 3m_1^2$, the allowed region is determined only by condition III. 
  At the point $A$, all the conditions become equivalent.}}
  \label{fig:MuConditions}
\end{figure}

These constraints on the parameter $\mu^4$ are plotted in Fig.~\ref{fig:MuConditions}. The unshaded regions correspond to values of $\mu^4$ consistent 
with a physical spontaneous symmetry-breaking phase, satisfying all of the previously mentioned conditions. The various constraints on $\mu^4$ can be summarised as follows:
\begin{itemize}
\item If $m_2^2 < \frac{m_1^2}{3}$ then $\mu^4 < m_2^4$ (Condition II);
\item If $\frac{m_1^2}{3} < m_2^2 < m_1^2$ then $\mu^4 < T_h$ (Condition III);
\item If $m_1^2 < m_2^2 < 3m_1^2$ then $\mu^4 < T_h$ (Condition III) or $T_H < \mu^4 < m_1^2 m_2^2$ (Conditions I and III);
\item If $3m_1^2 < m_2^2$ then $\mu^4 < m_1^2 m_2^2$ (Condition I).
\end{itemize}

\subsection{Equations of motion after SSB}

After expressing the full Lagrangian in terms of fluctuations around the vevs as done in Eq. (\ref{eq:fluctuation}), 
we can express the equations of motion after symmetry breaking in terms of the gauge fields $Z^\alpha$, $W^\alpha$ and $A^\alpha$. Introducing the notations
\be
C_+^\alpha\equiv\frac{J_{+, 1}^\alpha - i J_{+,2}^\alpha}{\sqrt{2}}~,\qquad K^\alpha_+ \equiv J^\alpha_{+,3} \cos \theta_{\rm W} - I^\alpha_{+} \sin \theta_{\rm W}~,
\ee
and
\begin{gather}
\sigma\equiv \mathbb{I} + \tau_3~,\qquad \omega \equiv \frac{\tau_3 \cos^2\theta_{\rm W} - \sin^2\theta_{\rm W}}{\cos \theta_{\rm W}}~,\nonumber 
\\ \tau_+\equiv\frac{\tau_1 - i \tau_2}{\sqrt{2}}~,\qquad \tau_-\equiv\frac{\tau_1 + i \tau_2}{\sqrt{2}}~,
\end{gather}
the equations of motion read as follows:

\noindent {\it Scalar fields}
\begin{subequations}
\begin{align}
0 &= D_\alpha D^\alpha \hat{\Phi}_1 + D_\alpha \left( \frac{ig}{2} Z^\alpha \omega + \frac{ig}{2} W^\alpha \tau_-  \right) V_1 - \frac{\mu^4}{m_2^2} \hat{\Phi}_1 \nn
&\phantom{=\ }+ \frac{\kappa}{2} \left( V_1^\dagger \hat{\Phi}_1 + \hat{\Phi}_1^\dagger V_1 \right)V_1 + \mu^2 \hat{\Phi}_2 
+ \frac{\kappa}{2} |\hat{\Phi}_1|^2 V_1+ \frac{\kappa}{2} \left( V_1^\dagger \hat{\Phi}_1 + \hat{\Phi}_1^\dagger V_1 \right) \hat{\Phi}_1 + \frac{\kappa}{2} |\hat{\Phi}_1|^2 \hat{\Phi}_1~,\\
0 &= D_\alpha D^\alpha \hat{\Phi}_2 + D_\alpha \left( \frac{ig}{2} Z^\alpha \omega + \frac{ig}{2} W^\alpha \tau_-   \right) V_2 + m_2^2 \hat{\Phi}_2 - \mu^2 \hat{\Phi}_1~;
\end{align}
\end{subequations}

\noindent {\it $Z^\alpha$ gauge field}
\bea
0 &=& \partial_\beta Z^{\alpha \beta} + ig \cos \theta_{\rm W} \left( W_\beta^\dagger W^{\beta \alpha} -  W^{\dagger \beta \alpha}W_\beta \right) 
+ \frac{1}{\xi} \partial^\alpha \partial^\beta Z_\beta \nn 
&&+ \frac{g^2 }{2 \cos^2\theta_{\rm W}} \left( |V_1|^2 + |V_2|^2 \right) Z^\alpha - K^\alpha_+ + ig \cos \theta_{\rm W} \left( \partial^\alpha \bar{\chi}^\dagger \chi  
- \partial^\alpha \bar{\chi} \chi^\dagger \right)
\nn
&&+ \frac{g^2}{2}  Z^\alpha \sum_i \Big( \hat{\Phi}_i^\dagger \omega^2 \hat{\Phi}_i + \left[ V_i^\dagger \hat{\Phi}_i
+ \hat{\Phi}_i^\dagger V_i \right] \Big) + \frac{eg}{2} A^\alpha \sum_i \Phi_i^\dagger \left( \omega \sigma \right) \hat{\Phi}_i 
\nn
&&- \frac{g g'}{2}\sin \theta_{\rm W}  \sum_i \left( \left[ \hat{\Phi}_i^\dagger \tau_- V_i \right] W^\alpha + \left[ V_i^\dagger \tau_+ \hat{\Phi}_i \right] W^{\alpha \dagger} \right) 
\nn
&&- \frac{gg'}{2} \sin \theta_{\rm W} \sum_i \left( \left[ \hat{\Phi}_i^\dagger \tau_- \hat{\Phi}_i \right] W^\alpha + \left[ \hat{\Phi}_i^\dagger \tau_+ \hat{\Phi}_i \right] W^{\alpha \dagger} \right)~;
\eea

\noindent {\it $A^\alpha$ gauge field}
\bea
0 &=& \partial_\beta F^{\alpha \beta} + ig \sin \theta_{\rm W} \left( W_\beta^\dagger W^{\beta \alpha} - W^{\dagger \beta \alpha} W_\beta \right) + \frac{1}{\xi} \partial^\alpha \partial^\beta A_\beta 
\nn 
&&- Q^\alpha + ig \sin \theta_{\rm W} \left( \partial^\alpha \bar{\chi}^\dagger \chi  - \partial^\alpha \bar{\chi} \chi^\dagger \right) \nn
&&+ e^2 A^\alpha \sum_i \hat{\Phi}_i^\dagger \sigma \hat{\
\Phi}_i + \frac{eg}{2}Z^\alpha \sum_i \hat{\Phi}_i^\dagger \left( \omega \sigma \right) \hat{\Phi}_i \nn
&&+ \frac{eg}{2} \sum_i 
\left( 
\left[\hat{\Phi}_i^\dagger \tau_- \hat{\Phi}_i \right] W^\alpha + 
\left[ \hat{\Phi}_i^\dagger \tau_+ \hat{\Phi}_i \right] W^{\alpha \dagger} + \left[ \hat{\Phi}_i^\dagger  \tau_- V_i \right] W^{\alpha} + \left[ V_i^\dagger \tau_+ \hat{\Phi}_i \right]W^{\alpha \dagger}
\right)~;
\eea

\noindent {\it $W^\alpha$ gauge fields}
\bea
0 &=& \partial_\beta W^{\alpha \beta} + ig W_\beta \left( \sin \theta_{\rm W} F^{\beta \alpha} + \cos \theta_{\rm W} Z^{\beta \alpha} \right) 
+ \frac{1}{\xi} \partial^\alpha \partial^\beta W_\beta \nn &&- ig \left( \sin \theta_{\rm W} A_\beta + \cos \theta_{\rm W} Z_\beta \right) W^{\beta \alpha} \nn 
&&+ \frac{g^2}{2} W^\alpha \left( |V_1|^2 + |V_2|^2 \right) - C_+^\alpha + ig \left( \partial^\alpha \bar{\chi} \eta_3 - \partial^\alpha \bar{\eta_3} \chi \right) 
 \nn 
&&+ \frac{g^2}{2} W^\alpha \sum_i \left( V_i \hat{\Phi}_i +  \hat{\Phi}_i^\dagger V_i + |\hat{\Phi}_i|^2  \right) \nn 
&&+ \frac{g}{2} \left( e A^\alpha - g' \sin \theta_{\rm W} Z^\alpha \right) \sum_i \left( \hat{\Phi}_i \tau_+ \hat{\Phi}_i + V_i^\dagger \tau_+  \hat{\Phi}_i \right)~.
\eea
From these equations, we can see that the gauge field masses are 
\be
M_W = g \sqrt{\frac{v_1^2 + v_2^2}{2}}=\cos \theta_{\rm W}M_Z\qquad \text{and}\qquad M_A = 0~,
\ee
as in the Hermitian Standard Model.

\subsection{Comments on the exceptional points}
\label{sec:exceptionalpoints}

At the zero exceptional points $\mu^2=\pm m_2^2$, the vevs become
\bea
v_1^2=v_2^2\equiv v^2=\frac{2}{\kappa}(m_1^2-m_2^2)~,
\eea
which vanish in the degenerate limit $m_1^2=m_2^2$. For $m_1^2 \neq m_2^2$, though, the gauge boson masses at the exceptional points are
\begin{equation}
M_W^2=g^2v^2=\cos^2 \theta_{\rm W}M_Z^2\neq 0~,
\end{equation}
remaining physical and non-zero. 

In order to make sense of this, in spite of the divergence of the $\mathcal{PT}$ norm and the apparent non-normalisability of the Goldstone modes (see Subsection~\ref{sec:vevs}),
it is helpful to reconsider the behaviour of the non-Hermitian theory at the exceptional point.  
As an example, let us consider the following $2 \times  2$ squared mass matrix of the non-interacting theory \cite{alexandre2017symmetries}: 
\be
\mathbf{M}^2=\left(\begin{array}{c c} m_1^2 & \mu^2 \\ -\mu^2 & m_2^2\end{array}\right)~.
\ee
For $m_1^2>m_2^2$, the eigenvectors of this squared mass matrix are
\be
\mathbf{e}_+=N\left(\begin{array}{c} \eta \\ \sqrt{1-\eta^2}-1\end{array}\right)\qquad \text{and}\qquad \mathbf{e}_-
=N\left(\begin{array}{c} 1-\sqrt{1-\eta^2} \\ -\eta\end{array}\right)~,
\ee
where
\be
\eta=\frac{2\mu^2}{m_1^2-m_2^2}~
\ee
(not to be confused with the ghost field appearing earlier).
The eigenvectors are not orthogonal with respect to the usual Hermitian inner product:
\be
	\mathbf{e}^{\ast}_+\cdot \mathbf{e}_-=2N^2\eta(1-\sqrt{1-\eta^2})~,
\ee
except in the Hermitian limit $\mu\to 0$ ($\eta \to 0$). They are, however, orthogonal with respect to the $\mathcal{PT}$ inner product, and orthonormality fixes
\be
N=\left(2\eta^2-2+2\sqrt{1-\eta^2}\right)^{-1/2}~.
\ee

The exceptional point of this mass matrix occurs when $\eta\to 1$, at which point the normalisation of the eigenvectors diverges. 
This signals that the mass matrix has become defective, having the Jordan normal form
\be
\label{eq:Jordanform}
\left.M^2\right|_{\eta\to 1}=\left(\begin{array}{c c} (m_1^2+m_2^2)/2 & 1 \\ 0 & (m_1^2+m_2^2)/2\end{array}\right)~,
\ee
and we lose an eigenvector. In fact, we see that  in the limit $\eta\to 1$ the eigenvectors $\mathbf{e}_+$ and $\mathbf{e}_-$ become parallel to one another. 
However, the issue of the non-orthogonality of these eigenvectors is then moot, and we can normalise them with respect to the Hermitian inner product, fixing
\be
\left.N\right|_{\eta=1}=\frac{1}{\sqrt{2}}~.
\ee
In other words, at the exceptional point the system behaves like a Hermitian theory with one fewer degree of freedom.

Returning to the case of spontaneously-broken gauge symmetries at the zero exceptional point, 
the explanation for the non-vanishing masses of the gauge bosons is 
that the Goldstone modes must be normalised with respect to Hermitian conjugation and not $\mathcal{PT}$ conjugation (which has become ill-defined). 
The discontinuity in the behaviour of the system as we approach exceptional points means that we must treat 
separately these particular points in parameter space.

Thus, our conclusion is that it is possible to give masses to gauge bosons in a gauge-invariant way through SSB also for non-Hermitian theories, 
even at the exceptional points. At these points, however, the counting of eigendirections must allow for the fact that the Hamiltonian has become defective. 

We note that different results were derived in
Ref.~\cite{Mannheim:2018dur}, which is based on an alternative interpretation of a similar (Abelian) version of this non-Hermitian theory, 
and where the gauge boson masses are zero at the zero exceptional point. The difference in our results can be traced back to differing interpretations of the complex conjugate: 
we take complex conjugation to act linearly on the fields, whereas in Ref.~\cite{Mannheim:2018dur} it is taken to act antilinearly on one of the fields (as motivated by a similarity transformation to a Hermitian theory). 
This has the effect of interchanging $v_2^2\to-v_2^2$ in the expression for the gauge boson masses, such that they then vanish at the zero exceptional point, when $v_1^2=v_2^2$. 
It is then argued that this is consistent with the fact that the Goldstone modes cannot be normalised with respect to the 
$\mathcal{PT}$ norm, which diverges at exceptional points, 
and these modes therefore cannot be ``eaten'' by the gauge field. This then leads Ref.~\cite{Mannheim:2018dur} to conclude that it is possible to break the gauge symmetry of a 
non-Hermitian model spontaneously without giving a mass to the gauge bosons. Our conclusion is the opposite: the gauge boson remains massive in the symmetry-broken phase, even at 
the zero exceptional point.

\section{Masses in the Non-Hermitian Model Compared with the Hermitian Model}

\begin{figure}[h]
  \includegraphics[width=1\linewidth]{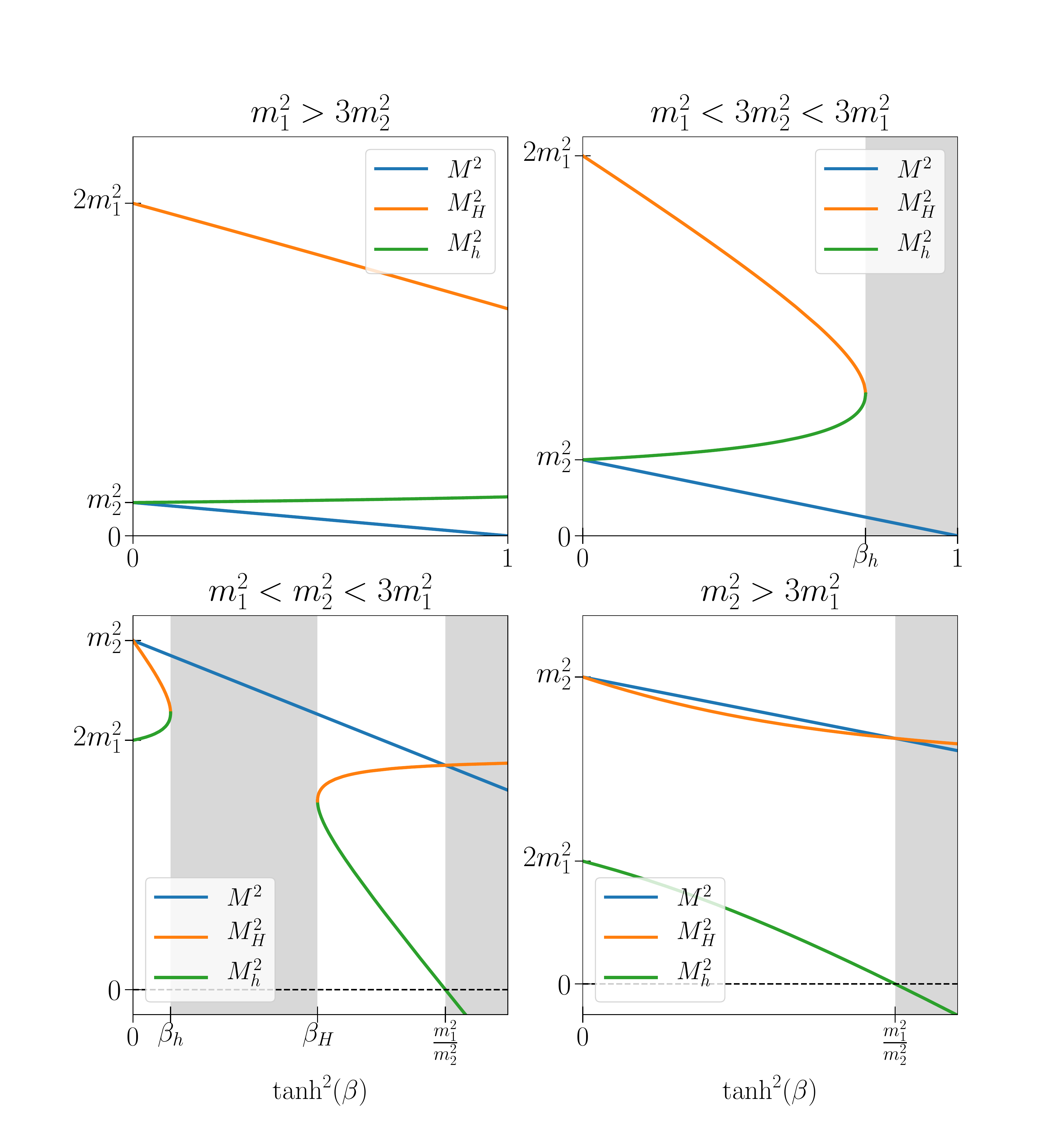}
  \caption{\it The masses of the physical scalar bosons as functions of $\tanh^2 \beta$ in different parameter regions.
Unphysical parameter regions are shaded grey.
 The upper left panel shows the region where $m_1^2 > 3 m_2^2$, 
 the upper right panel shows the region where $m_1^2 < 3 m_2^2 < 3 m_1^2$, the lower left panel
 shows the region where $m_1^2 < m_2^2 < 3 m_1^2$, and the lower right panel shows the
 region where $m_2^2 > 3 m_1^2$.}  
  \label{fig:ScalarMasses}
\end{figure}

\begin{figure}[h]
  \includegraphics[width=1\linewidth]{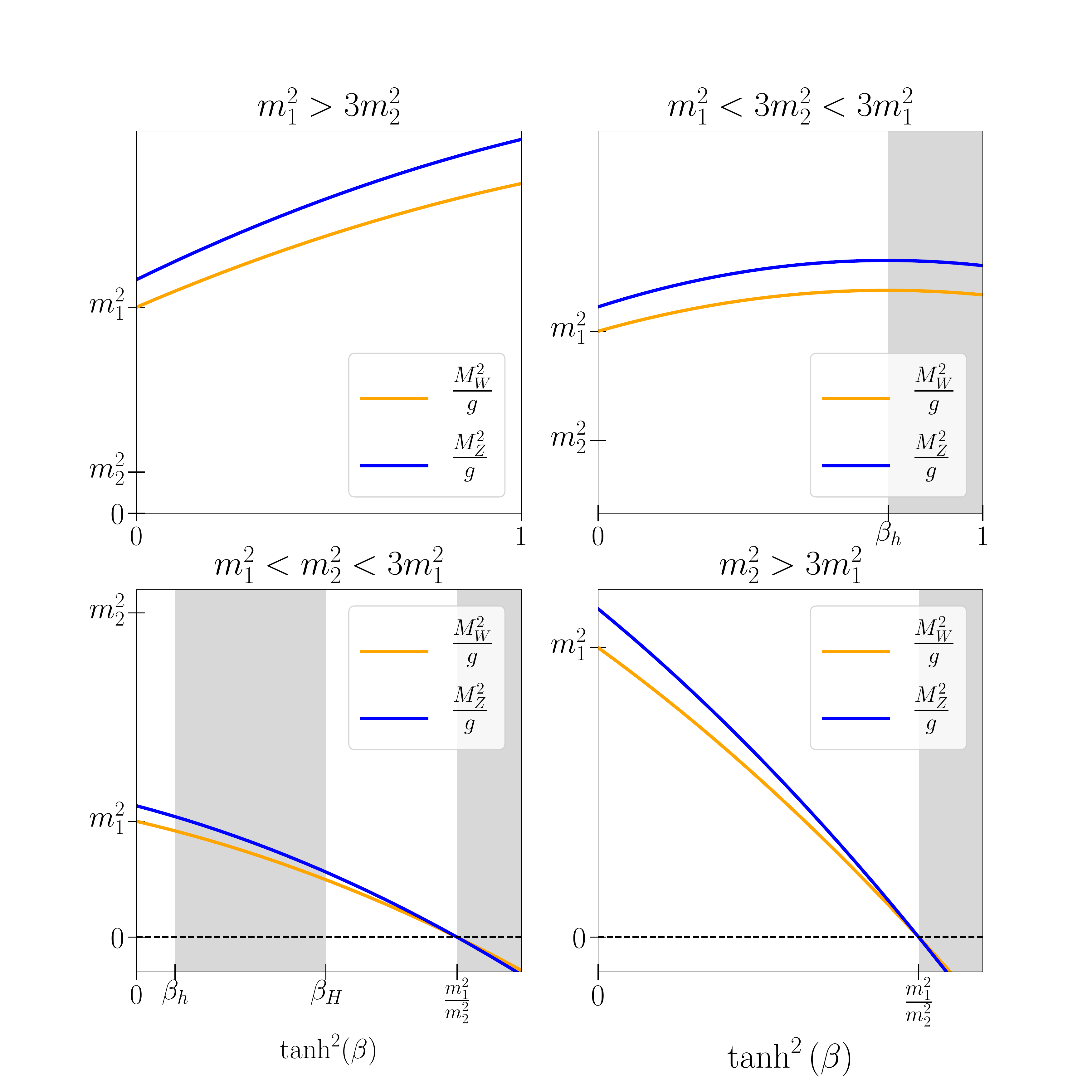}
  \caption{\it The masses of the charged and neutral gauge bosons as functions of $\tanh^2 \beta$ in the same parameter regions as in Fig.~\protect\ref{fig:ScalarMasses}.
  Unphysical parameter regions are shaded grey.}
    \label{fig:GaugeMasses}
\end{figure}

In this Section, we discuss the dependences of the
scalar and vector masses in the non-Hermitian 2HDM on the non-Hermitian mixing parameter $\mu^2$. These dependences are shown in Figs.~\ref{fig:ScalarMasses} and \ref{fig:GaugeMasses} for the scalar and vector bosons, respectively, wherein we have introduced the notation $\beta_{H(h)} \equiv T_{H(h)}/m_2^4$. In addition, we make a comparison with the dependence of the scalar and vector masses on a Hermitian mixing parameter in the corresponding Hermitian 2HDM.

We note the following features from each panel of Fig.~\ref{fig:ScalarMasses}:
\begin{itemize}

\item  In the region $m_1^2 > 3 m_2^2$, the mass $M^2$ goes to zero at the exceptional point $\mu^2=m_2^2$. If $\mu^2$ were to become larger then $m_2^2$
then $M^2$ would become negative and we would enter the phase of broken $\mathcal{PT}$ symmetry. 

\item In the region $m_1^2/3 < m_2^2 < m_1^2$, the masses $M_H^2$ and $M_h^2$ become equal at the point $\tanh^2 \beta = \beta_h$. For larger values of $\mu^2$, 
both  $M_H^2$ and $M_h^2$ would become complex. 

\item For $m_1^2 < m_2^2 < 3 m_1^2$, the masses $M_H^2$ and $M_h^2$ become equal at the point $\tanh^2 \beta = \beta_h$ or $\tanh^2 \beta = \beta_H$. 
Between these points, $M_H^2$ and $M_h^2$ become complex. When $\tanh^2 \beta > m_1^2/m_2^2$, the mass $M_H^2$ becomes negative. The unshaded regions correspond to physical masses.

\item For $m_2^2 > 3 m_1^2$, the masses are all real and positive as long as $\tanh^2 \beta < m_1^2/m_2^2$. 

\end{itemize}
We note in the lower right panel of Fig.~\ref{fig:GaugeMasses} that the gauge-boson masses vanish at the point $\mu^4 = m_1^2 m_2^2$, where the symmetry is restored, as we would expect.
  
It is interesting to compare the masses in this $\mathcal{PT}$-symmetric
non-Hermitian model with those in a similar Hermitian 2HDM with the following Lagrangian, involving a Hermitian mass-mixing term:
\bea \label{eq:HermScalarLagrangian}
\mathcal{L} &=& \partial_\alpha \Phi_1^\dagger \partial^\alpha \Phi_1 + \partial_\alpha \Phi_2^\dagger \partial^\alpha \Phi_2 + m_1^2 |\Phi_1|^2 
- m_2^2 |\Phi_2|^2 \nn &&+ m_{12}^2 \left( \Phi_1^\dagger \Phi_2 + \Phi_2^\dagger \Phi_1 \right) - \frac{\kappa}{4}|\Phi_1|^4~.
\eea
The vacuum expectation values for this Lagrangian are
\be
\langle \Phi_1 \rangle = \begin{pmatrix}
0 \\ v_1^H
\end{pmatrix} = V_1^H~,\qquad \langle \Phi_2 \rangle = \begin{pmatrix}
0 \\ v_2^H
\end{pmatrix} = V_2^H~, 
\ee
with
\be
v_1^H = \sqrt{\frac{2}{\kappa} \left( m_1^2 + \frac{m_{12}^4}{m_2^2} \right)}~,\qquad 
v_2^H = \frac{m_{12}^2}{m_2^2}\sqrt{\frac{2}{\kappa} \left( m_1^2 + \frac{m_{12}^4}{m_2^2} \right)}~.
\ee
After expressing the Lagrangian in terms of the shifted field $\hat{\Phi}_i$ where
\begin{subequations}
\bea
\Phi_i = \hat{\Phi}_i + V^H_i = \begin{pmatrix}
\phi_i^+ \\ v_i^H + \rho_i + i \psi_i
\end{pmatrix}~,\\
\Phi^\ast_i = \hat{\Phi}_i^* + V^H_i = \begin{pmatrix}
\phi_i^- \\ v_i^H + \rho_i - i \psi_i
\end{pmatrix}~,
\eea
\end{subequations}
we can calculate the eigenvalues. As in the non-Hermitian model, the massless states consist
of massless charged scalar and pseudoscalar Goldstone fields
\begin{subequations}
\bea
G^\pm &=& \frac{1}{\sqrt{(v_1^{H})^2 + (v_2^H)^2}} \left( v_1^H \phi_1^\pm + v_2^H \phi_2^\pm \right)~,\\
G &=& \frac{1}{\sqrt{(v_1^H)^2 + (v_2^H)^2}} \left( v_1^H \psi_1 + v_2^H \psi_2 \right)~.
\eea
\end{subequations}
The normalisations of the eigenmodes should be compared with those in Subsection~\ref{sec:vevs}. We remark that this Hermitian model is not $\mathcal{PT}$ 
symmetric if $\Phi_1$ and $\Phi_2$ transform as a scalar and a pseudoscalar, respectively. It is, however, $\mathcal{PT}$ symmetric if both $\Phi_1$ and $\Phi_2$ 
transform as scalars or pseudoscalars, and the Hermitian and $\mathcal{PT}$ norms coincide, as is expected for a Hermitian, $\mathcal{PT}$-symmetric theory.

\begin{figure}[h]
  \includegraphics[width=1\linewidth]{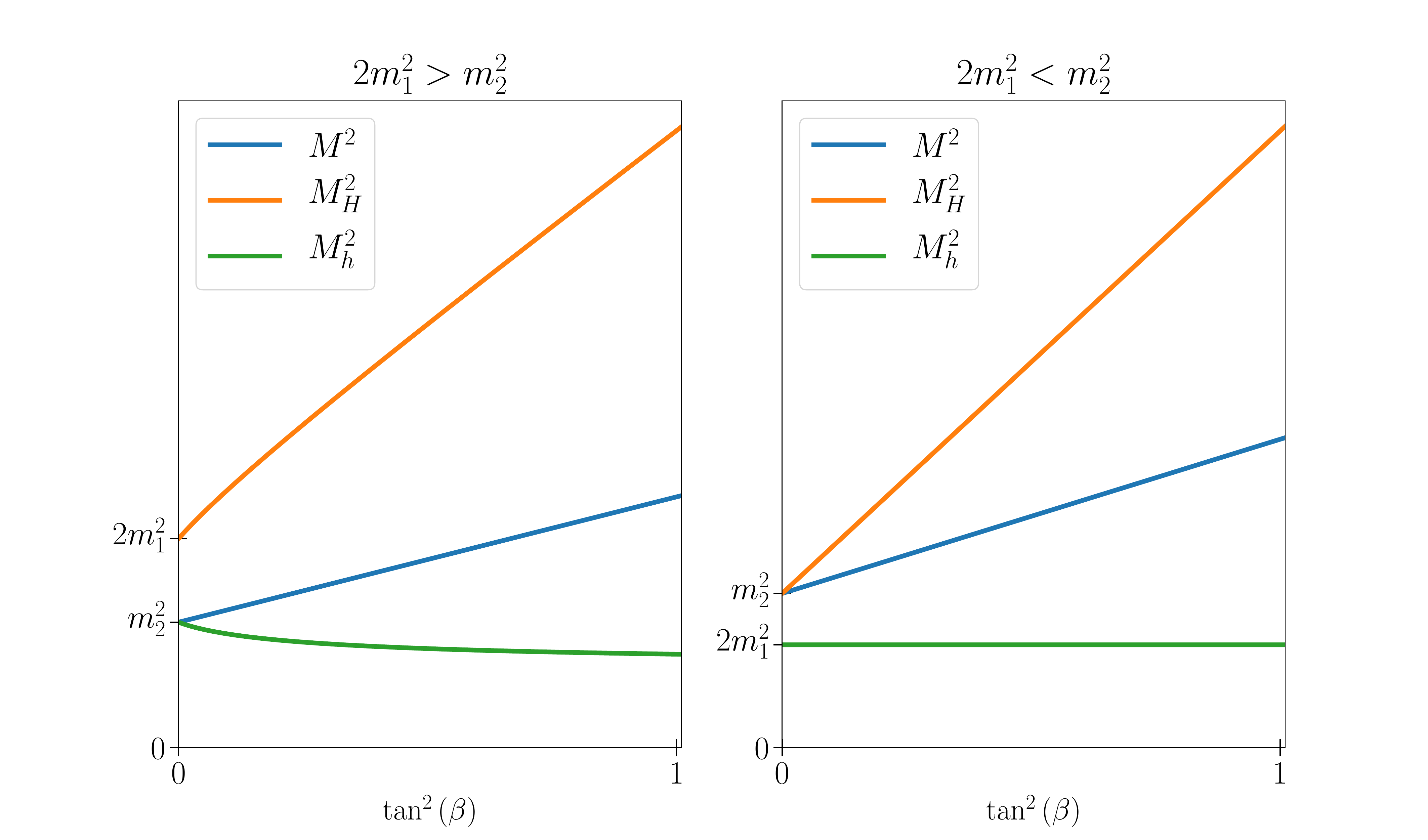}
  \caption{\it The masses of the scalar fields in the Hermitian two-Higgs-doublet model
  as functions of $\tan^2 \beta$ in the parameter ranges $2 m_1^2 > m_2^2$ (left panel) and $2 m_1^2 < m_2^2$ (right panel).}
    \label{fig:HScalarMasses}
\end{figure}

The remaining massive fields include a charged scalar, a neutral pseudoscalar and two neutral scalar fields. 
The charged scalars are
\be 
H^\pm = \frac{1}{\sqrt{(v_1^H)^2 + (v_2^H)^2}} \left( v_2^H \phi_1^\pm - v_1^H \phi_2^\pm \right)~,
\ee 
and the pseudoscalar is 
\be 
D = \frac{1}{\sqrt{(v_1^H)^2 + (v_2^H)^2}} \left( v_2^H \psi_1 - v_1^H \psi_2 \right)~,
\ee
with degenerate squared mass
\be
M^2 = \frac{(v_1^H)^2 + (v_2^H)^2}{v_1^H v_2^H} m_{12}^2~.
\ee Lastly, we can express the neutral scalar boson fields as 
\begin{subequations}
\bea
H &=& - \rho_1 \cos \alpha - \rho_2 \sin \alpha~,\\ h &=& \rho_1 \sin \alpha - \rho_2 \cos \alpha~,
\eea
\end{subequations}
with squared masses
\begin{subequations}
\bea
M_h^2 &=& \left( (v_1^H)^2 + (v_2^H)^2 \right) \left[ \lambda - \frac{\hat{\lambda} \cos \left( \beta - \alpha \right)}{\sin \left( \beta - \alpha \right)} \right]~,\\
M_H^2 &=& \left( (v_1^H)^2 + (v_2^H)^2 \right) \left[ \lambda + \frac{\hat{\lambda} \sin \left( \beta - \alpha \right)}{\cos \left( \beta - \alpha \right)} \right]~,
\eea
\end{subequations}
where
\begin{subequations}
\bea 
\tan \alpha &=& \frac{-m_{12}^2}{\left( M^2_H - m_2^2 \right)}~,\\
\tan \beta &=& \frac{v_2^H}{v_1^H}~,
\eea
\end{subequations}
and
\begin{subequations}
\bea 
\lambda &=& \kappa\,\cos^4  \beta ~,\\
\hat{\lambda} &=& \frac{\kappa}{2} \sin  2 \beta\, \cos^2  \beta ~.
\eea
\end{subequations}
The squared masses for this Hermitian model are plotted in Fig.~\ref{fig:HScalarMasses}
in the parameter ranges $2 m_1^2 > m_2^2$ (left panel) and $2 m_1^2 < m_2^2$ (right panel). 
We see that the mass spectra are completely different from the non-Hermitian, $\mathcal{PT}$-symmetric
case, offering distinctive phenomenological possibilities.

Before concluding, we remark that, by comparing the expressions above with those in Subsection~\ref{sec:vevs}, we can see that the 
non-Hermitian 2HDM that we have considered in this work is an analytic continuation of the Hermitian 2HDM, obtained by taking $m_{12}^4\to -\mu^4$. 
In other words, the Hermitian 2HDM lies in the fourth quadrant of the $(m_2^2/m_1^2,\mu^4/m_1^4)$ plane, not shown in Fig.~\ref{fig:MuConditions}.

\section{Conclusion}

In this paper, we have exhibited a consistent description of a non-Abelian two-Higgs-doublet model
with a non-Hermitian scalar mass-mixing term, which 
generalises the non-Hermitian extension of the Abelian Higgs model given in Refs.~\cite{AEMS1,AEMS2}. 
As in Ref.~\cite{AEMS2}, the main point that leads 
to a consistent model in the present article consists of restricting gauge invariance to a sub-class of gauge field configurations. The corresponding constraint plays the 
role of a conventional gauge-fixing condition, but which must be taken into account at the 
classical level already, in order to find consistent field equations. 
Within this framework, we have described the realisation of SSB and compared its features with the Hermitian case.

An interesting question is the significance of the exceptional points. As explained in this article, the number of eigendirections is reduced there,
so that this limit is not continuous. It is indeed easy to see that, in the non-interacting model, one can write a unique equation of motion for $\Phi_1+\Phi_2$ only, with mass $(m_1^2+m_2^2)/2$, when taking 
the exceptional limit $|\mu^2|\to |m_1^2-m_2^2|/2$, cf.~Eq.~\eqref{eq:Jordanform}. 
The introduction of gauge {or self-}interactions does not allow this though,
and one can therefore question the stability of the exceptional points under quantum corrections, 
which appear as soon as interactions are switched on. However, the treatment of radiative corrections
and further study of the exceptional points goes beyond the scope of the present article.

We have noted that physical observables depend on $\mu^4$, 
and thus not on the set of equations of motion we choose.
This can be checked also with the masses of scalar excitations and gauge bosons: 
the transformation $\mu^2\to-\mu^2$ leads to changes in the signs of $\alpha$ and $\beta$ [see Eq.~(\ref{alphabeta})],
such that the masses obtained after SSB are not modified. It was shown in Ref.~\cite{AEMS2} that the quantum theory also depends on $\mu^4$ only, and we expect the same to be valid here, since
this feature is based on the scalar sector properties of the partition function, which is very similar here. 

Finally, we note that the scalar boson mass spectrum in the non-Abelian non-Hermitian model differs
significantly from that in the Hermitian version. This shows that the non-Hermitian model opens up new
phenomenological perspectives, which merit a subsequent more detailed discussion.

\section*{Acknowledgements}

The authors thank Andreas Fring, Philip Mannheim and Takanobu Taira for interesting discussions. 
The work of JA, JE and DS was supported by the United Kingdom STFC Grant ST/P000258/1, and that of JE also by the Estonian Research Council via a Mobilitas Pluss grant. 
The work of PM was supported by a Leverhulme Trust Research Leadership Award (Grant No.~RL-2016-028).

\end{document}